\documentstyle[11pt,newpasp,twoside,psfig]{article}
\def\edcomment#1{\iffalse\marginpar{\raggedright\sl#1\/}\else\relax\fi}
\reversemarginpar

\begin{document}
\title{The Dawn of Galaxies: Deep MAMBO Imaging Surveys} 
\author{F. Bertoldi, K.M. Menten, E. Kreysa}
\affil{Max-Planck-Institut f\"ur Radioastronomie, 53121 Bonn, Germany}
\author{C.L. Carilli, F. Owen}
\affil{NRAO, Socorro, NM 87801, USA}

\begin{abstract}
We discuss results from sensitive, wide-field imaging of the 
millimeter extragalactic background using the Max-Planck Millimeter
Bolometer array (MAMBO) at the IRAM 30~m telescope.
\end{abstract}


The detection of a far-infrared and sub-millimeter wavelength background
by COBE, and the first sensitive, high-resolution SCUBA images of the
sub-mm background have caused significant revisions of our picture of
the star-formation history of the universe. SCUBA at 850 $\mu$m
discovered a population of what appear to be star forming galaxies at
high redshifts, most of which are invisible at optical and near-IR
wavelengths (Smail et al. 1997; Hughes et al. 1998; Barger et al. 1999;
Eales et al. 1999; Bertoldi et al. 2000).  The infrared
luminosities of these objects are comparable to the bolometric
luminosities of QSOs, but their optical faintness shows that, unlike for
QSOs, nearly all of the bolometric luminosity arises from thermal
emission of dust grains. The emitting dust is probably heated in
massive, optically obscured star forming regions, with star formation
rates $\sim 10^3\rm M_\odot yr^{-1}$.  The objects
discovered with SCUBA and MAMBO 
can account for the integrated (sub)mm
background, and thereby about 25$\%$ of the total 
infrared background radiation. The relation between metal production
and the extragalactic background (Eales et al. 1999)
would then imply that at least a quarter of all stars
were formed during the extreme starbursts we now see as thermal
background sources.

The total number of (sub)mm background sources discovered with
SCUBA at 850 $\mu$m (350 GHz), and with the 37-channel MAMBO array at
1.2~mm (250 GHz) now 
exceeds 100, and great efforts have been made trying to identify
the class of known objects to which  these (sub)mm sources
belong. 
However, most of them defy a clear optical identification, and thereby
an accurate redshift determination.  The positional accuracies of the
SCUBA and MAMBO sources are of order 5$''$, which for each source 
allows for a number of faint optical and
near-IR sources as possible counterparts, objects 
which  are usually too
faint for spectroscopic studies. To date, only two MAMBO sources have 
clear near-IR identifications, and three SCUBA sources have clear
optical or near-IR identifications with a spectroscopic redshift (Lilly
et al. 1999; 
Ivison et al. 2000).

More promising are recent attempts to identify the (sub)mm background
sources with radio sources in deep VLA images. We found that the majority
of several dozen brightest and highest-S/N MAMBO sources in
our 100~arcmin$^2$ map of the Abell 2125 region have 20~cm radio
counterparts within 5$''$ of the MAMBO source position.  Since in our
$7.5 \mu$Jy rms noise 20~cm VLA map 
on average 
we find only one $>5\sigma$ source per
arcmin$^2$, chance alignments of radio and MAMBO sources are rare, and
nearly all of the MAMBO-radio associations should be real.  The radio
identification of a MAMBO source determines its position within an
arcsecond, and thereby allows a unique identification of any optical or
near-IR counterpart, or as in most cases, the lack of one.

\begin{figure}[]
\plotfiddle{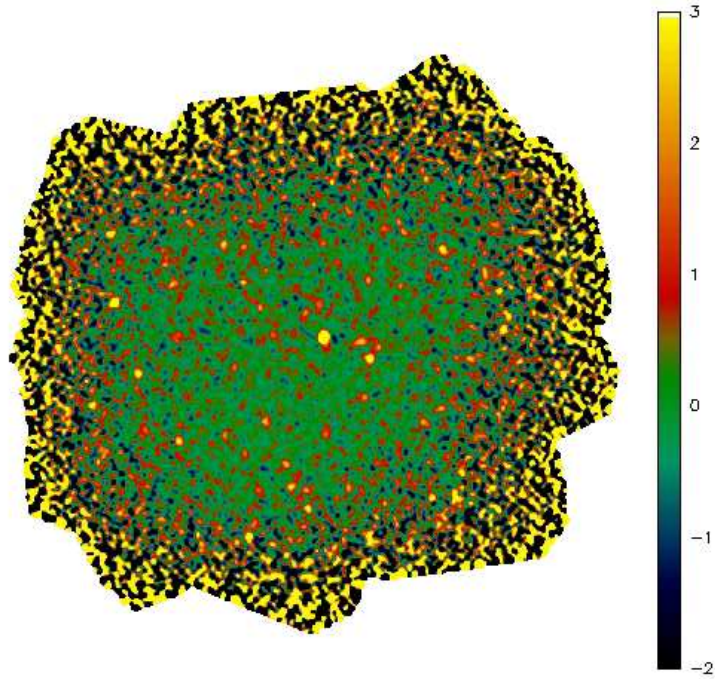}{8.5cm}{-90}{60}{60}{-170}{310}
\caption{A preliminary  $\approx 14$ arcmin
  diameter 1.2~mm MAMBO image of the NTT Deep 
  Field. The rms noise is 0.5~mJy in the image center, and
  rises toward the edges. 
  At least 20 significant sources can be seen, including
  the strong $z=4.7$ quasar BR 1202-0725.  During the last two winters
  we used MAMBO to obtain maps of three fields with a total area of
  about 300~arcmin$^2$ covered to a noise rms $<1$ mJy.  
  We found almost 100 point sources
  with fluxes above $1.5$~mJy (Bertoldi et al.  2000; Carilli et
  al. 2000; Menten \& Bertoldi 2000).
}
\end{figure}

{\bf A radio identification also allows an approximate redshift
determination.} Relying on the tight correlation between the
radio and the far-IR flux densities of star forming galaxies, Carilli \&
Yun (1999) showed that the radio-to-(sub)mm flux ratio decreases with
increasing redshift to such 
an extent that the observed value of this flux ratio can be used to
estimate redshifts up to $z\approx 4$. 
Although contributions to the
radio flux from an AGN, or systematically higher dust temperatures
compared to local starburst galaxies would lead to an underestimate of
the redshifts (Blain 1999), the radio-to-(sub)mm flux ratios of several
dozen SCUBA and MAMBO sources provide a unique first look at their
redshift distribution, placing most of them at $z\approx 1$ to 4
(Fig.~2). Including SCUBA/MAMBO sources for which upper limits to their
radio fluxes are known, the redshift distribution shifts to slightly
higher values, but it appears unlikely that there exists a dominant
population of SCUBA/MAMBO sources at very high ($z>4$) redshifts.


\begin{figure}[h]
\begin{minipage}{9cm}
\psfig{figure=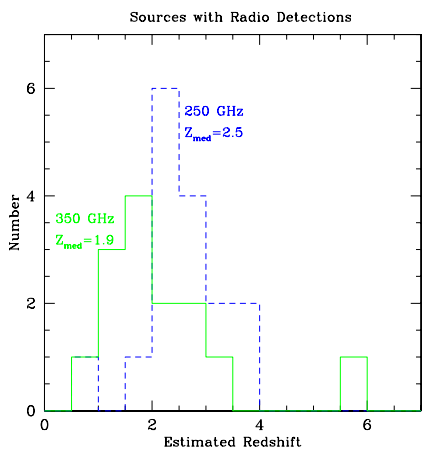,width=2.35in}
\end{minipage}
\begin{minipage}{8cm}
\caption{ Redshift distribution 
  for 250 GHz (1.2~mm) and 350 GHz (850 $\mu$m) selected sources,
  based on the cm-to-mm flux density ratio for sources with
  radio detections at 1.4 GHz,
  including redshift lower limits based on radio upper limits.}
\end{minipage}
\end{figure}

Figure 3 shows the preliminary cumulative source counts based on two of
our three MAMBO deep fields, along with source counts determined from
various SCUBA surveys. We relate the 250 GHz flux densities to 350 GHz
flux densities using a scaling factor of 2.25, applicable to a typical
starburst galaxy at $z \approx 2.5$. We have included faint source
counts in the regions within a 1$'$ radius of the cluster center
assuming a mean gravitational magnification factor of 2.5.
The MAMBO and SCUBA counts agree well at intermediate flux densities,
and they show a steepening of the distribution at $S_{350} \approx 10$ mJy.

\begin{figure}[ht]
\begin{minipage}{9cm}
\psfig{figure=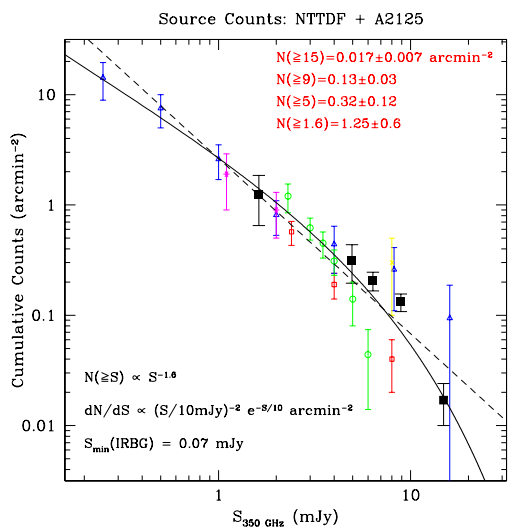,width=7cm,angle=0,clip=t}
\end{minipage}
\begin{minipage}{7cm}
\caption{Preliminary cumulative source counts from two MAMBO fields 
  as large solid squares, plus counts from various SCUBA surveys.  The
  dashed curve is a power law of index --1.6.  All of the data can be
  reasonably fit by an integrated Schechter-type luminosity function,
  with parameters as given on the plot.}
\end{minipage}
\end{figure}


{\bf What is the possible cause of a high brightness turnover
of the (sub)mm background source counts?} 
If these objects are indeed starbursts, then
the implied star formation rates would be $>2000 \rm~ M_\odot yr^{-1}$.
Such extreme rates could only be sustained for some minimum 
(dynamical, free-fall, dissipation) timescale by the most massive
galaxies.  A turnover in the brightness distribution could thus
indicate a turnover in the mass function of galaxies, or alternatively,
it may be the signature of an upper limit to the luminosity of a
starburst, an equivalent to the Eddington limit, determined by the
energy output of the burst and the ability of the
gas to dissipate energy rapidly enough to permit further mass
infall or cloud collapse.

{\bf What is the nature of the (sub)mm background sources?}  It appears
unlikely that a large fraction of them are dust-enshrouded QSOs, since
Chandra observations failed to detect most of the targeted SCUBA sources
(e.g. Fabian et al. 2000).
However, two of the three SCUBA sources for which optical emission lines
were seen show signatures of an active nucleus.  
The brightest source found in the MAMBO blind survey
is an intermediate-redshift QSO, showing non-thermal
emission at mm wavelengths. In a SCUBA mapping survey Knudsen et
al.  (2000) also find their brightest object to be a
$z=2.8$ QSO. 
These two objects could be coincidental, but they do hint
at a possible overlap of the (sub)mm background and the QSO populations.

Most likely, much of the energy emitted by the (sub)mm background
sources arises from starbursts.  Because then they produce a significant
fraction of all stars in the Universe, and because they are more
luminous than any starburst galaxy in the local universe, it is
suggestive to think that they are elliptical galaxies seen at the time
when they formed most of their stars. It would be of great interest to
establish their exact redshift distribution. If they had formed over a
wide range of redshifts, they probably formed through the hierarchical
merging of smaller objects.
If they formed in a
narrow time interval at high redshift however, it would suggest that
they formed through the collapse of single, massive primordial density
enhancements. Our current estimate of their redshift distribution is
consistent with the latter picture, the monolithic collapse of
$10^{12}\rm M_\odot$ structures, but this remains to be verified.

\vspace{.4cm}

{\it \noindent The MAMBO surveys are a collaborative effort also involving
R. Zylka, L. Rei\-chertz, A. Bertarini, D. Lutz, H. Dannerbauer, 
L. Tacconi, and R. Genzel.}

\end{document}